\documentclass{iaus}
\usepackage{graphicx}
\usepackage{amsmath,amssymb}
\usepackage{natbib}

\title[Massive stellar models] 
{Massive stellar models:\\ rotational evolution, metallicity effects}

\author[Ekstr\"om et al.]   
{Sylvia Ekstr\"om$^1$,
 Cyril Georgy$^1$,
 Georges Meynet$^1$,\\
 Andr\'e Maeder$^1$,
 \and Anah\'i Granada$^{1,2}$}

\affiliation{$^1$Geneva Observatory, University of Geneva \\ Maillettes 51 - Sauverny,
CH-1290 Versoix, Switzerland \\[\affilskip]
$^2$Instituto de Astrof\'isica de La Plata, Universidad Nacional de La Plata, \\ Paseo del Bosque S/N, La Plata, Buenos Aires, Argentina}

\pubyear{2010}
\volume{272}  
\jname{Active OB stars - structure, evolution, mass loss, and critical limits}
\editors{C. Neiner, G. Wade, G. Meynet \& G. Peters, eds.}

\begin{document}

\maketitle

\begin{abstract}
The Be star phenomenon is related to fast rotation, although the cause of this fast rotation is not yet clearly established. The basic effects of fast rotation on the stellar structure are reviewed: oblateness, mixing, anisotropic winds. The processes governing the evolution of the equatorial velocity of a single star (transport mechanisms and mass loss) are presented, as well as their metallicity dependence. The theoretical results are compared to observations of B and Be stars in the Galaxy and the Magellanic Clouds.
\end{abstract}

\firstsection 
\section{Introduction}
Why care about rotation? Just because stars do rotate! A look at the velocity distribution established by \citet{HG06a} reveals that the peak in the probability density occurs at $\upsilon_\text{eq}\simeq200$ km/s, which represents a ratio $\upsilon/\upsilon_\text{crit}\simeq0.5-0.6$, \textsl{i.e.} a substantial fraction of the keplerian velocity.

The effects of rotation in stars were studied since the works of von Zeipel and Eddington in the years 1924-1925. In the end of the 60s, they were included in polytropic or simplified stellar models \citep{roxburgh65,RoxbStritt66,faulkner68,KippTh70,EndSof76}. About 30 years later, stellar models became more sophisticated and also benefited from the inclusion of rotational effects \citep{pinson89,Deupree90,FliegLang94,chab95,Meynet96}.

Since the end of the 90s, more or less extended grids of rotating stellar models were computed \citep{langer97,mm1,SiessLiv97,hlw00}. Those grids showed that the inclusion of the effects of rotation improved the adequation between models and massive stars observations in many aspects:
\begin{itemize}
\item the surface abundances of light elements \citep{HL00,mm5};
\item the predicted surface velocities in clusters \citep{mart06,mhff06,mm5};
\item the blue- to red-supergiants number ratio in the SMC \citep{mm7};
\item the WR populations number with metallicity \citep{mm10,mm11,vdk05};
\item the rotation rates of pulsars \citep[when strong coupling is assumed,][]{hws05};
\item the SN types and GRB progenitors \citep{mm11, YLN06,georgy09}.
\end{itemize}

Roughly summarised, rotating stars are expected to present
\begin{enumerate}
\item a modified gravity:
  \begin{itemize}
  \item the surface characteristics become dependent on the colatitude considered;
  \item there is a mass loss enhancement and anisotropy;
  \end{itemize}
\item chemical species and angular momentum transport mechanisms:
  \begin{itemize}
  \item the behaviour on the HR diagram is modified;
  \item the nucleosynthesis is altered;
  \item there is a surface enrichment;
  \item the mass loss is modified;
  \item the rotation profile evolves during the star life, becoming steeper (when no magnetic fields are considered).
  \end{itemize}
\end{enumerate}

We will go through the different points in this review.
\section{Surface characteristics}
\subsection{Gravity and shape}
Because of rotation, the effective gravity is modified, becoming a function of the rotation velocity $\Omega$ and of the colatitude $\theta$ of the star:
\begin{align*}
\vec{g}_\text{eff}=\vec{g}_\text{eff}(\Omega,\theta)=&\left( -\frac{GM}{r^2}+\Omega^2r\,\sin^2\theta \right)\,\vec{e}_r\\
&+ \Omega^2r\,\sin\theta\,\cos\theta\,\vec{e}_\theta.
\end{align*}
We immediately see that at the pole ($\theta=0^o$) the effective gravity is just the gravitation acceleration $-GM/r^2$. At the equator ($\theta=90^o$) the centrifugal force adds a sustaining term $\Omega^2r\,\sin^2\theta$. In these two cases, the effective gravity is still radial, while at intermediate $\theta$, the term $\Omega^2r\,\sin\theta\,\cos\theta$ does not vanish and implies that the effective gravity is no more radial.

In the frame of the Roche model, the maximal oblateness allowed when the star rotates at the critical velocity\footnote{The critical velocity is reached when $\vec{g}_\text{eff}$ vanishes because the centrifugal force at the equator counterbalances the gravity exactly (see Section~\ref{sec_critlim}).} is $R_\text{eq,crit}=1.5\,R_\text{pol,crit}$.

Recently, interferometry has allowed to determine the deformation of fast rotating stars. A first evaluation of the oblateness of Achernar by \citet{domi03} showed a larger ratio $R_\text{eq}/R_\text{pol}$ than the one allowed in the Roche model. However, more recent observations of the same star \citep{vinicius06,carciof08} have revised this ratio and found a lower value, more compatible with these theoretical expectations.
\subsection{Flux and effective temperature}
Since the flux is related to the effective gravity \citep{vZ24,owocki96,owocki98,Maed99}, it becomes also dependent on colatitude:
$$
\vec{F}=\vec{F}(\Omega,\theta)\simeq-\frac{L}{4\pi\,G\,M^\star}\,\vec{g}_\text{eff}(\Omega,\theta)
$$
with $M^\star=M\,\left( 1-\frac{\Omega^2}{2\pi\,G\,\rho_m} \right)$ the so-called \textsl{reduced mass} which takes into account the reduction of the gravitational potential by rotation ($\rho_\text{m}$ is the mean density inside the considered isobar).

The Stefan-Boltzmann law $F=\sigma\,T^4$ implies a dependence on the colatitude for the effective temperature as well:
$$
T_\text{eff}=T_\text{eff}(\Omega,\theta)=\left[ \frac{L}{4\pi\sigma\,G\,M^\star}\,\vec{g}_\text{eff}(\Omega,\theta) \right]^{1/4}.
$$

\begin{figure}
\centering
\includegraphics[width=.6\textwidth]{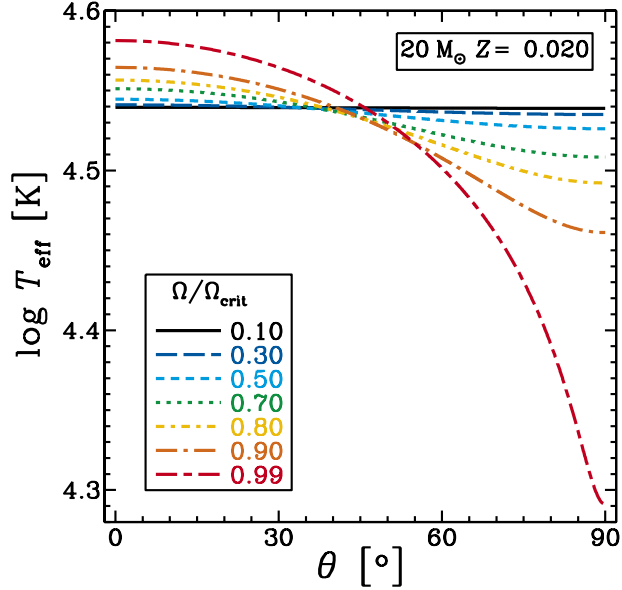}
\caption{Variation of the effective temperature with the colatitude for various initial rotation rates \citep[figure from][]{emmb08}.\label{f_teffcolat}}
\end{figure}

Recent interferometric observations, like for example \citet{monnier07} on Altair or \citet{zhao09} on $\alpha$ Cephei and $\alpha$ Ophiuchi, have provided a possibility to test this relation. Let's have a close look at Altair: the rotation rate $\Omega/\Omega_\text{crit}$ of this star is evaluated between 0.90 and 0.92, and the temperature difference between the pole and the equator is between 1.19 and 1.32. For such a rotation rate, the theoretical models (cf. Fig.~\ref{f_teffcolat}) predict a difference of 1.26-1.33, in good agreement with the observational value.
\subsection{Mass loss}
According to the works of \citet{OwGay97}, \citet{mm6} and \citet{PP2000}, rotation enhances the mass loss by a factor:
$$
\frac{\dot{M}(\Omega)}{\dot{M}(0)}=\left[ \frac{\left( 1-\Gamma_\text{Edd} \right)}{\left( 1-\frac{\Omega^2}{2\pi\,G\,\rho_m}-\Gamma_\text{Edd} \right)} \right]^{\frac{1}{\alpha}-1}
$$
where $\Gamma_\text{Edd}$ is the Eddington factor, \textsl{i.e.} the ratio of the luminosity of the star to the Eddington luminosity $L_\text{Edd}=\frac{\kappa_\text{s}}{4\pi\,c\,G\,M}$, with $\kappa_\text{s}$ the electron-scattering opacity.

It also changes the geometry of the mass flux \citep{Maed02,MaederBook09}, which is no longer constant on the whole surface of the star. The mass loss by surface unit at a given latitude $\theta$ follows the relation:
$$
\frac{\text{d}\dot{M}(\theta)}{\text{d}\sigma} \sim A(\alpha,k)\ \left(\frac{L}{4\pi\ G\ M^\star}\right)^{\frac{1}{\alpha}-\frac{1}{8}}\ \frac{g_\text{eff}(\theta)^{1-\frac{1}{8}}}{\left( 1-\Gamma_\Omega(\theta) \right)^{\frac{1}{\alpha}-1}}
$$
where $A(\alpha,k)=(k\alpha)^{1/\alpha}\left( \frac{1-\alpha}{\alpha} \right)^{(1-\alpha)/\alpha}$ is a function of the force multiplier parameters which characterise the stellar opacity.

If we consider that $A=\text{cst}$, a star rotating for example at $\Omega/\Omega_\text{crit}=0.95$ will have $\dot{M}(\text{pol})=3.2\,\dot{M}(\text{eq})$. However, if one accounts for the change of the force multiplier parameters, \textsl{i.e.} a variation in the opacity regime caused by the drop of $T_\text{eff}$ at the equator, the equatorial mass loss can be enhanced at a given point, driving the formation of a decretion disc \citep{Ow04}.

On the observations side, interferometry again sheds a new light in this topic, showing features that can be interpreted as polar enhanced winds \citep{kerv06,meil07}, as well as discs around active stars \citep{meil07,schaefer2010}.

The geometry of the wind may leave an imprint on the circumstellar medium. There are some indications of asymmetry detected in spectropolarimetry observations of some supernovae \citep[see for example SN 2007rt by][]{trundle09}. This aspect is actually under study with 2- and 3D simulations and will be the subject of a future paper (Walder et al. in prep).
\section{Rotational evolution}
\subsection{Two competing processes}
The evolution of the surface velocity of a star is the result of the competition between two processes:
\begin{enumerate}
\item the \textbf{mass loss}, which removes angular momentum at the surface and thus decelerates the rotational velocity;
\item the \textbf{transport} of angular momentum inside the star, which brings some internal angular momentum to the surface (mainly through the \textsl{meridional circulation}) and may counterbalance the loss by the winds.
\end{enumerate}
\begin{figure}
\resizebox{\hsize}{!}{\includegraphics{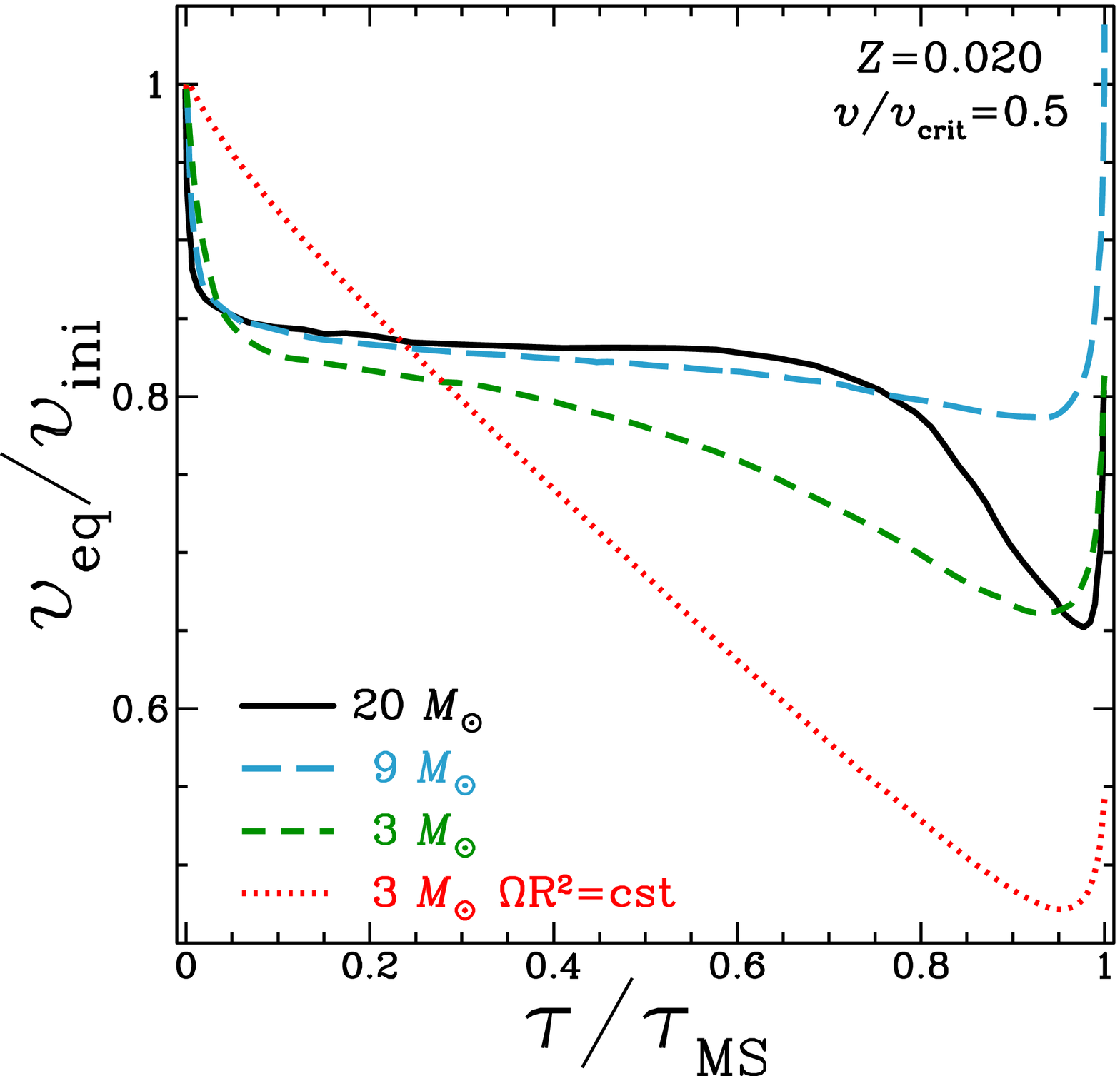}\hspace{.5cm}\includegraphics{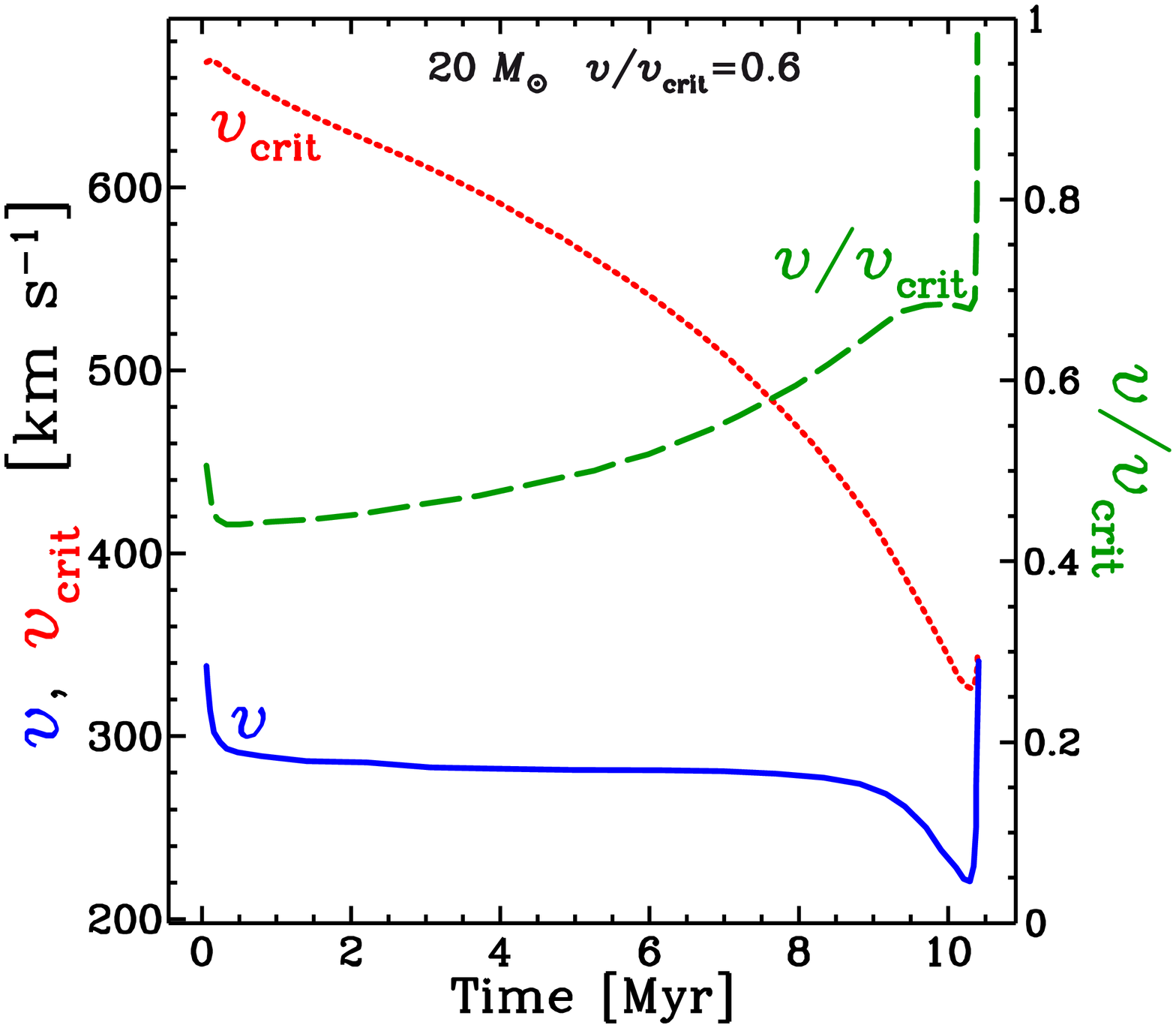}}
\caption{\textsl{Left:} evolution of the equatorial velocity (normalised to the initial velocity) during the main sequence for different mass domains. \textsl{Right:} evolution of the equatorial velocity, the critical velocity and the ratio $\upsilon/\upsilon_\text{crit}$ during the main sequence.\label{f_vevol}}
\end{figure}
Both processes are dependent on the mass of the star: the more massive the star, the stronger winds and larger meridional currents it experiences (see Fig.~\ref{f_vevol}, \textsl{left}). At solar metallicity typically, around 20 $M_\odot$ and above, the winds contribution wins so the star decelerates during the main sequence (MS). Around 9 $M_\odot$, both processes are in equilibrium so the surface velocity remains almost constant during the largest part of the MS. Around 3 $M_\odot$, there are almost no mass loss through winds, but the core-envelope coupling through the meridional circulation is so weak that the surface velocity decreases during the MS. Note that recent observations by \citet[][see also Poster S1-05 in these proceedings]{huang10} show a quicker deceleration of the less massive of the B stars than predicted by current theoretical models, more compatible with a regime of local angular momentum conservation $\Omega\,R^2=\text{cst}$.
\subsection{Critical limit\label{sec_critlim}}
\begin{figure}
\resizebox{\hsize}{!}{\includegraphics{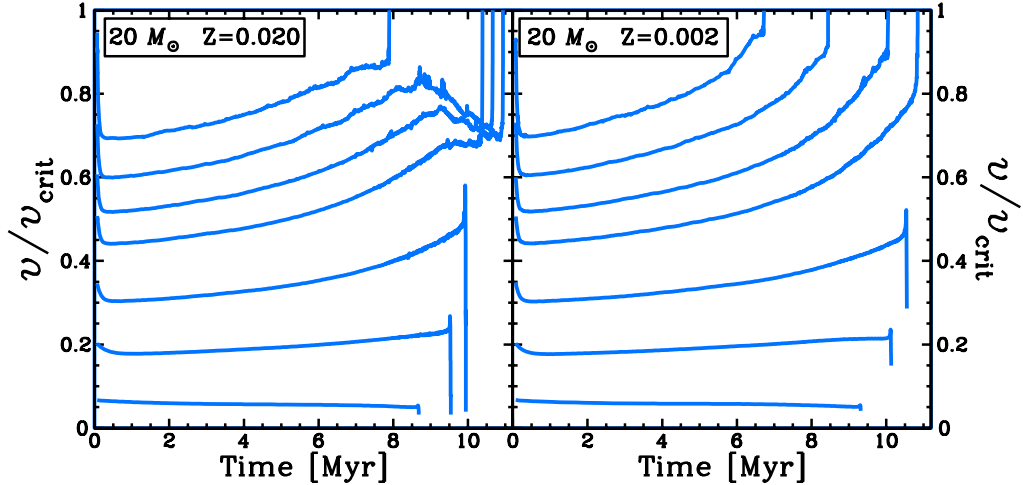}}
\caption{Evolution of the $\upsilon/\upsilon_\text{crit}$ ratio during the main sequence for 20 $M_\odot$ models at $Z=0.020$  and $Z=0.002$.\label{f_vvcevol20}}
\end{figure}
One thing is the evolution of the surface velocity, another thing is the evolution of the ratio to the critical limit $\upsilon_\text{crit}=\sqrt{\frac{2}{3}\frac{G\,M}{R_\text{pol}}}$. During the MS, the mass may be reduced by mass loss mechanisms, and the radius steadily inflates, so the critical limit drops, as shown in Fig.~\ref{f_vevol}, \textsl{right}. The result is that although the surface may decelerates, the star may encounter the critical limit at a moment in its main sequence lifetime. Comparing the left panels of Fig.~\ref{f_vvcevol20} and \ref{f_vvcevol9}, we see that the conditions to reach the critical limit during the MS are met if the star has not a too high mass, and also if it is not a too slow rotator at birth. For each mass domain, there is a minimal initial $\upsilon/\upsilon_\text{crit}$ ratio allowing for the reaching of the critical limit.

Once at the critical limit, the star remains close to it. It experiences phases of mass ejection (slowing the surface below the critical value) followed by quiescent phases, during which it slowly re-accelerates toward the critical limit.

It is possible now for theoretical models to evaluate the mass ejected in the form of a disc (see Poster S1-03 by Georgy et al. and S1-06 by Krti\v cka et al. in these proceedings). This 'mechanical' mass loss seems to occur at a lower rate than the one that can be measured around Be stars \citep{rinehart99,Stee03}. However, in the models, the mechanical mass loss is averaged on a much longer timestep than the period of mass ejection observed in Be stars. The instantaneous mass-loss rate is expected to be higher than the average one given by the models.
\subsection{Metallicity effects}
\begin{figure}
\centering
\resizebox{\hsize}{!}{\includegraphics{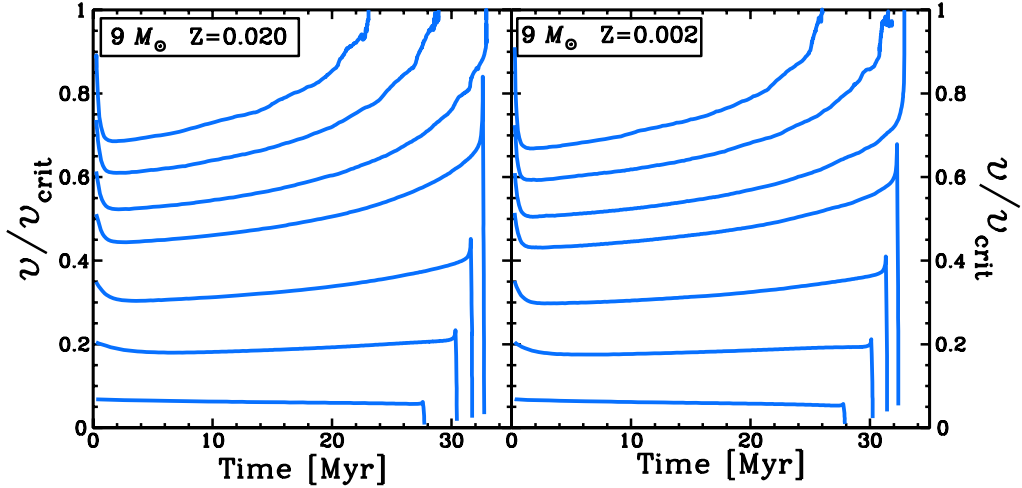}}
\caption{Evolution of the $\upsilon/\upsilon_\text{crit}$ ratio during the main sequence for 9 $M_\odot$ models at $Z=0.020$  and $Z=0.002$.\label{f_vvcevol9}}
\end{figure}
Both the radiative mass loss rate and the strength of the meridional circulation depend on the metallicity. A low metallicity reduces both mechanisms: there are less metal lines to interact with the photons and form a radiatively driven wind and the greater compactness of the star reduces the amplitude of the meridional circulation.

The net effect on the evolution of the equatorial velocity depends strongly on the mass domain considered: for the most massive stars, the reduction of the winds is the dominant effect, so the stars reach more easily the critical limit (see Fig.~\ref{f_vvcevol20}). On the contrary, for less massive stars for which the winds are not too strong anyway, the reduction of the meridional circulation is dominant, so the stars have more difficulties to reach the critical limit (see Fig.~\ref{f_vvcevol9}).
\subsection{Be phenomenon}
Fast rotation is supposed to be linked with the Be phenomenon. It is however not yet clear whether it may explain by itself the origin of the equatorial disc observed around Be stars. Observations show a ratio $\upsilon/\upsilon_\text{crit}\approx0.70-0.80$ \citep{Port96,chauv01,tyc05}. Maybe this ratio is true, but there may also be alternative explanations: first, as shown by \citet{town04}, there is a saturation effect in the widening of the lines by rotation, so the true velocity may be underestimated. Second, the Be phenomenon seems to consist of ejection phases followed by quiescent phases during which the stars may rotate with a lower rate.

Observations show a metallicity trend in the appearance of the Be phenomenon \citep[see for example \citealt{mgm99} or][]{WB06}. Two scenarios are evoked to explain the fast rotation of the Be stars: the binary channel \citep[see for example][]{McSG05b} (where the fast rotation arises from the accretion of angular momentum from the mass-donor companion), or the single star channel \citep{emmb08} (where the surface acceleration is a natural evolution due to the core-envelope coupling). Any such scenario should present this metallicity trend to be valid. To study the single star evolution scenario, \citet{emmb08} computed 112 stellar models (4 masses, 4 metallicities and 7 rotation rates) with the Geneva code. To get populations numbers, we convolved the models with a Salpeter IMF, using the velocity distribution of \citet{HG06a}.

A good agreement between the theoretical population ratio Be/(B+Be) and the observed one is obtained modulo two adjustments:
\begin{enumerate}
\item the Be phenomenon should appear already at 70\% of the critical velocity;
\item the velocity distribution should count more fast rotators at birth in low-metallicity environments.
\end{enumerate}
Point $(a)$ can be sustained by several mechanisms. For example, sub-surface convective motions have been shown to be able to give the needed impulse to eject matter from the surface \citep{mgm08,cantiello09}, as also would non-radial pulsations do \citep{mcswain08,Ow04}. The binarity of the star could play a role, here not by spinning-up the star but by adding a gravitational pull allowing to launch the matter in the disc \citep{kerv08}. Point $(b)$ presents contradictory observational supports. While \citet{mart07a} do find a higher mean $\Omega/\Omega_\text{crit}$ ratio in the SMC compared to the LMC and the Galaxy, \citet{PenGies09} draw opposite conclusions, finding no clear evidence for any difference in the velocity distribution at different metallicities.

Note that the stellar models have been computed without the effects of magnetic fields, the inclusion of which could change the picture. It may be that taking into account the strong core-envelope coupling brought by the magnetic fields relieves the theoretical prediction from the two aforementioned adjustments.
\subsection{LBVs}
Until now, the critical velocity we have considered was determined only by the centrifugal contribution $\vec{g}_\text{rot}$ against $\vec{g}_\text{grav}$ ($\Omega$-limit). However, a critical point appears whenever $\vec{g}_\text{tot}=\vec{g}_\text{grav}+\vec{g}_\text{rot}+\vec{g}_\text{rad}=0$. If $\vec{g}_\text{rot}=0$ or is negligible, the star may meet the classical Eddington-limit when $\vec{g}_\text{rad}=\vec{g}_\text{grav}$. If $\vec{g}_\text{rad}$ is negligible, the star may encounter the $\Omega$-limit when $\vec{g}_\text{rot}=\vec{g}_\text{grav}$: this describes the first critical velocity considered until now:
$$
\upsilon_\text{crit,1}=\sqrt{\frac{2}{3}\frac{GM}{R_\text{pol,crit}}}
$$
with $R_\text{pol,crit}$ the polar radius when the star is at the critical limit.

If all three terms are significant, the star may meet the $\Omega\Gamma$-limit when $\vec{g}_\text{tot}=0$. It means that for a given $\Omega$, there is a maximum luminosity given by:
$$
L_{\Gamma\Omega} = \frac{4\pi\,c\,G\ M}{\kappa} \left( 1-\frac{\Omega^2}{2\pi\,G\ \rho_\text{m}}  \right)
$$
with $\kappa$ the total opacity. Inversely, depending on the Eddington factor of the star, a second critical velocity can be defined:
$$
\upsilon_\text{crit,2}=\sqrt{\frac{9}{4}\,\upsilon_\text{crit,1}^2\frac{1-\Gamma_\text{max}}{V'(\omega)}\frac{R_\text{e}^2(\omega)}{R_\text{pol,crit}^2}}
$$
where the quantity $V'(\omega)=\frac{V(\omega)}{\frac{4\pi}{3}R^3_\text{pol,crit}}$ is the ratio of the actual volume of a star with rotation $\omega=\Omega/\Omega_\text{crit}$ to the volume of a sphere of radius $R_\text{pol,crit}$.

Note that for $\Gamma_\text{Edd} < 0.639$, $\upsilon_\text{crit,2}$ is not defined. Above this value, $\upsilon_\text{crit,2}$ becomes smaller than $\upsilon_\text{crit,1}$, so the star encounters the $\Omega\Gamma$-limit before the $\Omega$-limit. This could be the case of some known LBVs \citep{groh06,groh09b}, that present both a high Eddington factor and a high rotation rate. With a $\Gamma_\text{Edd}\simeq0.8$, they are probably meeting their second critical velocity already with $\Omega/\Omega_\text{crit}\simeq0.85-0.9$.
\section{Internal mixing}
\subsection{Diffusion of chemical species}
\begin{figure}
\centering
\resizebox{\hsize}{!}{\includegraphics{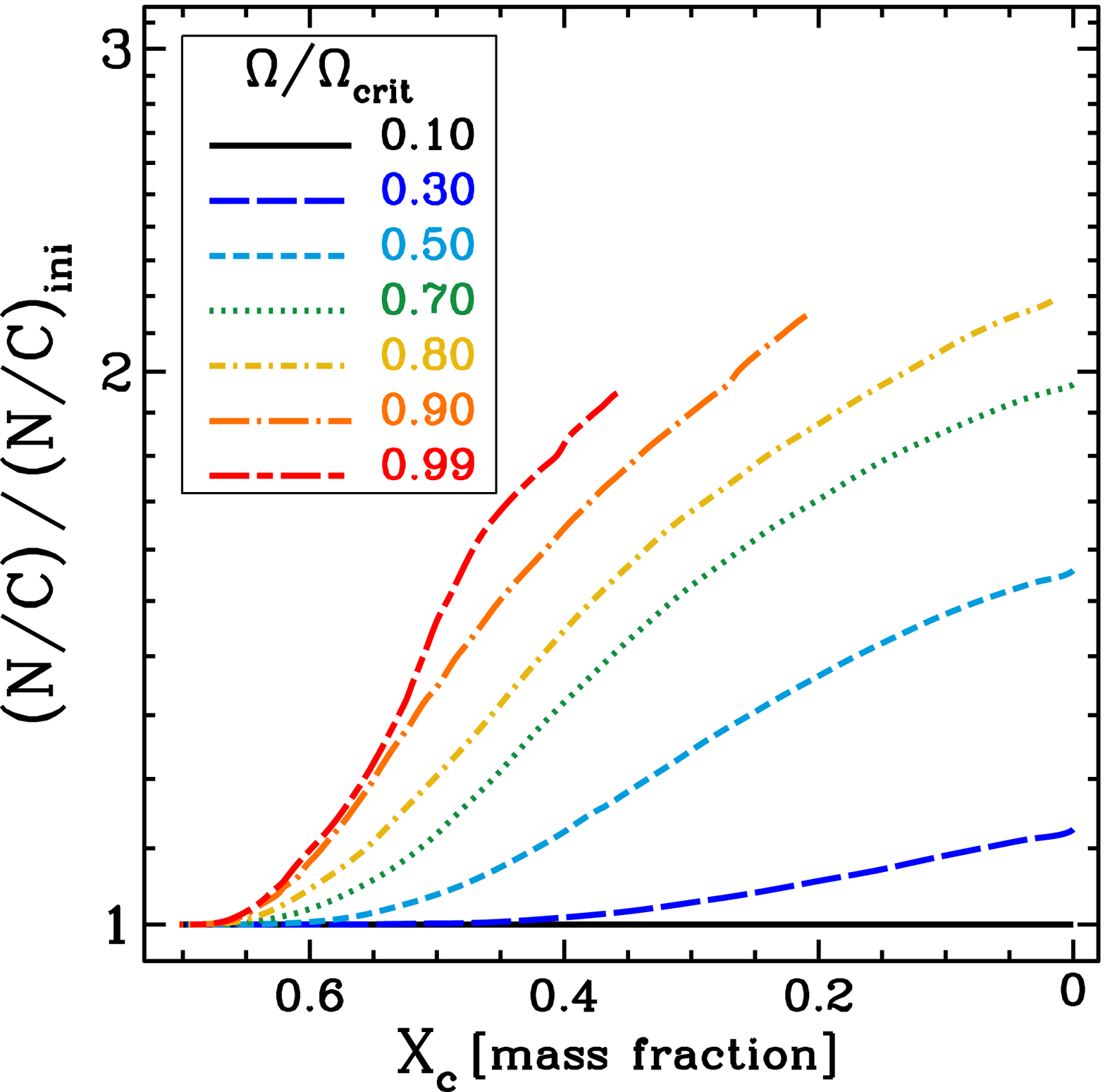}\hspace{.5cm}\includegraphics{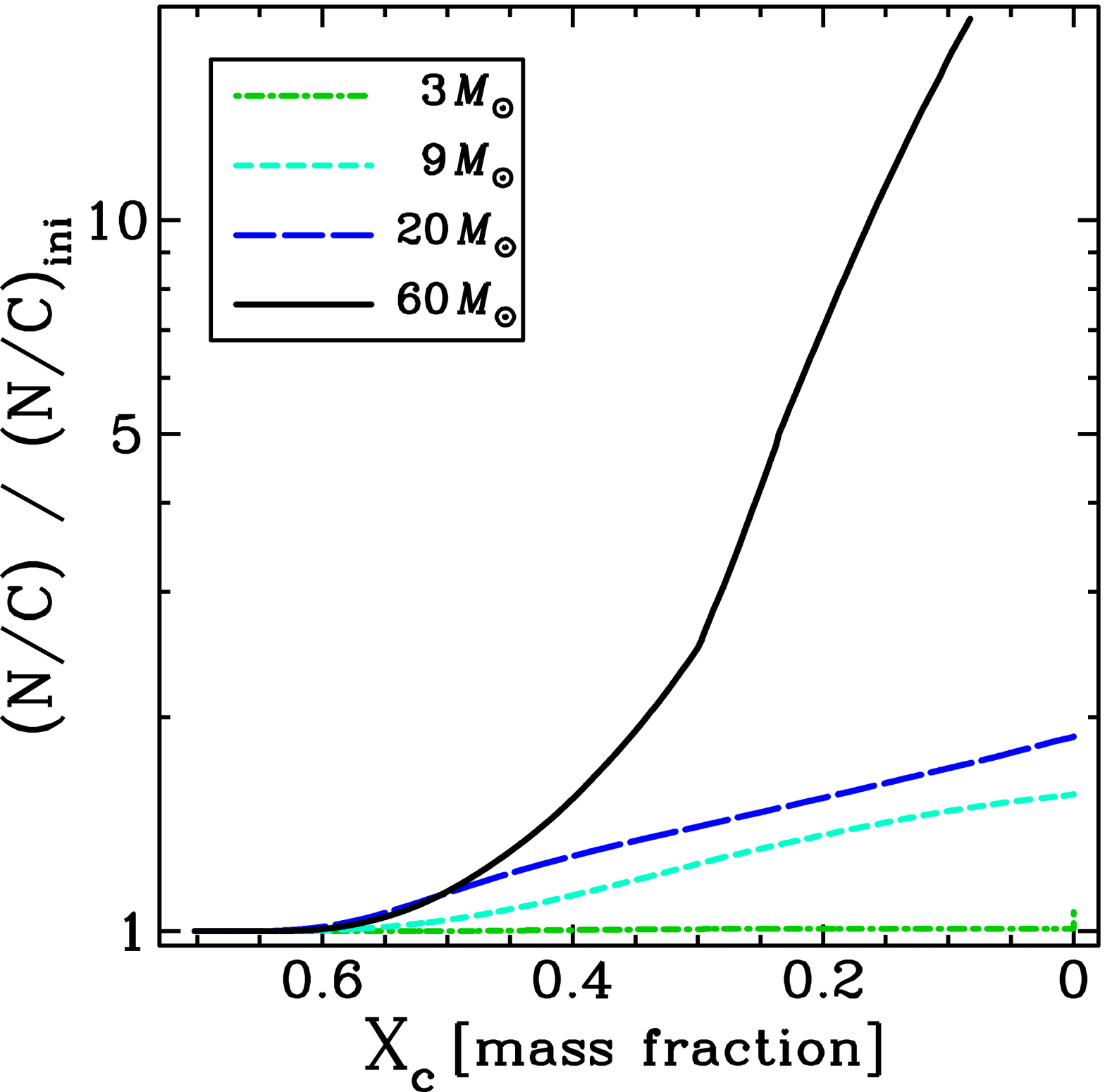}}
\caption{\textsl{Left:} evolution of the surface N/C ratio (normalised to the initial one) during the main sequence (expressed as the central mass fraction of hydrogen $X_\text{c}$) for 9 $M_\odot$ models at $Z=0.020$ with varying the initial ratio $\Omega/\Omega_\text{crit}$. \textsl{Right:} same as \textsl{left}, but for different mass domains.\label{f_ncevol}}
\end{figure}
The radiative zones inside a star are supposed to rotate differentially. This causes shear turbulence which in turn drives some turbulent mixing of chemical species. The higher the rotation rate, the stronger is the mixing (see Fig.~\ref{f_ncevol}, \textsl{left}).

We can evaluate the diffusion time: $\tau_\text{mix}\simeq\frac{R^2}{D}$ with $D$ the diffusion coefficient. With a relation between the radius and the mass, we get $\tau_\text{mix}\propto M^{-1.8}$ while the main sequence lifetime is $\tau_\text{MS}\propto M^{-0.7}$. We see that the more massive the star, the stronger is the mixing, leading to a surface enrichment already on the main sequence (see Fig.~\ref{f_ncevol}, \textsl{right}).
\subsection{Metallicity effects}
Low-metallicity stars are more compact than higher-metallicity ones. The meridional currents are less efficient, so the $\Omega$-profile inside the star is steeper. The mixing is thus stronger, while the mixing time is shorter, so we expect a strong surface enrichment on the main sequence.
\subsection{Surface abundances}
The surface abundances are expected to be modified by mixing. A good signature of mixing is the C, N, and O abundances: their respective ratios are expected to change, but their sum C+N+O is supposed to remain constant. Only at very low $Z$, primary nitrogen production may occur, leading to a net metallicity increase. A careful abundance analysis made by \citet{przybilla2010} in the solar neighbourhood shows that the observed mixing follows very well the trend expected from CNO nuclear reactions (see their contribution in the same proceedings). The adequation with stellar models is also good, but the observations are not very constraining yet.

It is very important to keep in mind that the mixing is a function of the rotation rate, of course, but also (as shown above) of the mass, the metallicity and the age, as well as other characteristics as the binarity for example. The study of the rotational mixing in stars should imply a separate analysis, in order to discriminate between several parameters. For example, while \citet{hunt08b}, studying field and clusters stars in the SMC, find many outliers from the expected path in the N/C vs $\upsilon_\text{rot}$, a re-analysis by \citet{maeder09} of a sub-group of the same stars, restrained to a narrow mass range and belonging to the same cluster (\textsl{i.e.} having the same age), shows that most of the outliers shifted back to the expected path. Most of the outliers left cannot be used to test the rotation-induced mixing, because they are evolved stars or known binaries. For the very few remaining, we indeed need another explanation than pure rotational mixing. There are several possibilities that shall be the subject of further studies, as the influence of magnetic fields inside the star or magnetic braking at the surface. However, the need for a yet unknown process is evoked (see the contribution of Ines Brott in the same proceedings).
\section{Conclusion}
Rotation is linked to many very interesting types of objects as Be stars or LBVs. The inclusion of its effects in theoretical stellar models is essential to understand those objects and their direct environment. While rotating models are in a better adequation with observations, there is still a lot of work to do in order to improve our understanding of the physical processes at work.

 Theoretical and numerical developments are continuously on-going, but theory alone is like a car without a driver. Observations are highly needed:
\begin{itemize}
\item \textsl{larger surveys} of well identified objects would allow to make a separate analysis of sub-groups in masses, ages, ..., with a significant statistics;
\item a larger number of objects studied by \textsl{interferometric measurements} would allow to put constraints on the surface characteristics of the models;
\item the progress in \textsl{asteroseismology} should provide constraints on the internal structure of the stars;
\item precise observations of the \textsl{circumstellar environment} could help to constrain the mass loss mechanisms, which are a key ingredient in massive stars.
\end{itemize}

\bibliographystyle{cup}
\bibliography{BibTexRefs}

\begin{discussion}

\discuss{S. Owocki}{I would like to emphasize an important difference between reaching the two critical rotation speeds identified in the Maeder \& Meynet analysis. The first critical speed applies to low luminosity stars like Be stars, and should lead to a circumstellar decretion disk that is ejected mechanically at the equator.  But for the second critical speed, which is modified by the radiative forces associated with $\Gamma > 0.65$, the mass loss should be mainly over the poles, not equator. In effect, the rapid rotation and associated equatorial gravity darkening forces the stellar luminosity to emerge over a smaller surface area over the poles, so that even if $\Gamma \lesssim 1$ the local flux over the poles can exceed the Eddington value, leading to a radiatively driven, bipolar, prolate mass loss. In short, in low luminosity, $\Gamma << 1$ Be stars, critical rotation leads to the observed equatorial disk, while in $\Gamma \lesssim 1$  LBVs it leads to bipolar nebulae.}

\discuss{O. Chesneau}{A comment to Stan: If the mass loss of LBVs is essentially prolate, and directed toward the pole, how can you explain why the environment of LBVs is dominated by ring-like structures (see comment of K. Weis)?}

\discuss{S. Owocki}{Well, in slowly rotating LBVs you should get spherical mass loss that will likely appear ring-like.  Perhaps Kerstin can comment on how relatively common these rings are vs. bipolar LBVs.}

\discuss{K. Weis}{As Stan already mentioned, LBVs create polar winds which lead to larger number of bipolar nebulae and not only spherical ring nebulae. See my talk on Thursday.}

\end{discussion}

\end{document}